\documentclass[a4paper]{jpconf}
\usepackage{graphicx}

\usepackage[labelformat=empty]{subcaption}
\usepackage{citesort}
\usepackage{microtype}

\begin{document}
\title{$S$- and $D$-wave vector charmonia}

\author{A Krassnigg$^{1,2}$ and T Hilger$^1$}

\address{$^1$Institute of Physics, University of Graz, NAWI Graz, A-8010 Graz, Austria}

\address{$^2$Institute of High Energy Physics, Austrian Academy of Sciences, A-1050 Vienna, Austria}

\ead{andreas.krassnigg@uni-graz.at}

%%%%%%%%%
\begin{abstract}
We revisit earlier calculations of leptonic decay constants of vector charmonia and present and illustrate our decomposition of the corresponding covariant Bethe-Salpeter amplitudes in terms of orbital angular momentum as interpreted in the meson's rest frame. Our results confirm our previous conclusions drawn from the magnitudes of vector-meson leptonic decay constants, identifying the  $\Psi(3770)$ and $\Psi(4160)$ as $D$-wave states in our setup.
\end{abstract}

%%%%%%%%%
\section{Fundamental approaches to hadron spectroscopy}
Hadrons have been around for a long time and, apparently, also will be \cite{Olive:2016xmw}. 
They have their prominent place in particle physics for a number of reasons. 
One of these reasons is presented by reference to the strong interaction, which, in the face of our understanding of the atomic nucleus as a composite object made of only  electrically positively charged and neutral nucleons, was necessary to provide a mechanism for the stability of matter as we know it.
Thus, it is important to keep in mind that it is the strong interaction that defines what a hadron is and not the hadron's inner workings.

As a consequence, it is a choice what is meant by a fundamental approach to hadron spectroscopy, simply because we can choose or invent different mechanisms to underly the hadron spectrum.
Many scientists also make many clever choices of this kind. 
A delectably large number of them were presented at least in some respect at the \textsc{Fairness2017} workshop, which should be understood as a merit of the organizing committee.
The landscape of such approaches has been impacted significantly by the quark concept and the advent of quantum chromodynamics (QCD) several decades ago.
Some people are tempted to rely on  the quark-gluon picture of hadrons too easily, e.\,g., when it comes to the distinction between mesons and baryons, which is actually and simply made by the value of the hadron's baryon number.
The fact that composing hadrons out of quarks and gluons reproduces the correct pattern is necessary as a feature of a successful explanation, but not acceptable as a definition.
Still, one can get the impression that, cf. \cite{Orwell:1945af}
\begin{quote}
All approaches to hadron spectroscopy are fundamental, but some approaches are more fundamental than others \ldots
\end{quote}
However, this notion, often referring to the relevant scales in different descriptions of hadrons, is sometimes unjustly used to imply truth or importance.
Instead, we understand the term \emph{fundamental approach} to hadron spectroscopy as one that explains the hadron spectrum by a well-defined set of rules and assumptions. 
We refer the reader to the other corresponding contributions to these proceedings for more details.

%%%%%%%%%
\section{DSBSE approach}\label{sec:dsbse}
Our approach to hadron spectroscopy is the Dyson-Schwinger-Bethe-Salpeter-equation (DSBSE) incarnation of QCD. 
It is fundamental at the level of quarks and gluons and starts at the QCD Lagrangian.
This framework of infinitely many coupled nonlinear integral equations is nonperturbative, manifestly covariant, and uses continuum quantum field theory \cite{Bashir:2012fs}.
In numerical studies, truncations are employed in order to render the computation of any given hadronic observable feasible. 
Investigation of schemes of truncations, e.g. \cite{Bhagwat:2004hn,Holl:2004qn,Gomez-Rocha:2014vsa,Sanchis-Alepuz:2015tha,Gomez-Rocha:2015qga,Gomez-Rocha:2016cji} and references therein, have pointed to the rainbow-ladder (RL) truncation \cite{Eichmann:2008ae} as a meaningful first step for any numerical treatment, in particular, if the set of results is to be comprehensive \cite{Krassnigg:2009zh,Popovici:2014pha,Hilger:2014nma,Hilger:2017jti}.
In addition, RL-truncated computations have always phenomenologically pioneered the DSBSE-approach, be it for the properties of mesons \cite{Maris:1999nt,Krassnigg:2003wy,Krassnigg:2003dr,Krassnigg:2004if,Holl:2004un,Alkofer:2005ug,Holl:2005vu,Cloet:2007pi,Eichmann:2008kk,Dorkin:2010pb,Blank:2010pa,Mader:2011zf,Raya:2015gva,Hilger:2016efh,Hilger:2016drj} or baryons \cite{Bloch:2003vn,Alkofer:2004yf,Holl:2005zi,Eichmann:2007nn,Nicmorus:2008vb,Nicmorus:2008eh,Eichmann:2008ef,Eichmann:2009qa,Eichmann:2009fb,Eichmann:2016yit}.

It rapidly becomes clear to any practitioner of theoretical hadron spectroscopy that dynamical chiral symmetry breaking (D$\chi$SB), an apparent phenomenon in QCD, is a necessary ingredient in a successful description of the light-hadron spectrum.
For example, D$\chi$SB provides a mass scale that, starting from current quarks with a mass of a few MeV each, results in the nucleons' mass of about $1$ GeV.
While constituent-quark masses can explain a lot in terms of the masses of  hadrons simply by adding them appropriately and thinking in terms of binding energies, the limits of this concept are illustrated best by investigation of the lightest hadron, the pion  \cite{Horn:2016rip}.

A better explanation is offered by symmetry arguments. The chiral symmetry of QCD with massless quarks, together with its dynamical breaking results in a massless Goldstone boson. 
The explicit breaking of QCD's chiral symmetry by the current-quark masses makes the pion the massive but very light realization of this (pseudo-)Goldstone boson. 
As a consequence, pion properties are an important focus of modern hadron theory.

\begin{table}[b]
\caption{Calculated pseudoscalar- and vector-meson decay constants $f_{\mathrm{Calc}}$ in MeV compared to experimental data for selected charmonia, together with our orbital-angular-momentum assignment $L_{\mathrm{Calc}}$ and contributions in percent.\label{tab:dconstants}}
\centering
\begin{tabular}{llrr||c|rrrrrr} \\
State & $J^{PC}$       & $f_{\mathrm{Exp}}$ \cite{Olive:2016xmw}  & $f_{\mathrm{Calc}}$ \cite{Krassnigg:2016hml} & $L_{\mathrm{Calc}}$ & $S$   & $S$-$P$ & $P$ & $P$-$D$ & $D$ & $S$-$D$	 \\ \hline\\[-2ex]
$\eta_c$ & $0^{-+}$    & $338(14)$  & $378$		 &   $S$      & $99$ & $1$ & $0$ & $-$ & $-$ & $-$  \\
$J/\Psi$ & $1^{--}$    & $416(6)$   & $411$	     &   $S$      & $99$ & $1$ & $0$ & $0$ & $0$ & $0$  \\
$\Psi(2S)$ & $1^{--}$  & $295(5)$   & $155$		 &   $S$      & $100$ & $0$ & $0$ & $0$ & $0$ & $0$  \\
$\Psi(3770)$ & $1^{--}$ & $100(4)$  & $45$		 &   $D$      & $0$ & $0$ & $0$ & $2$ & $98$ & $0$  \\
$\Psi(4040)$ & $1^{--}$ & $187(17)$ & $188$		 &   $S$      & $100$ & $0$ & $0$ & $0$ & $0$ & $0$  \\
$\Psi(4160)$ & $1^{--}$ & $143(32)$ & $1$		 &   $D$      & $0$ & $0$ & $0$ & $1$ & $99$ & $0$	\\
\end{tabular}\hfill
\end{table}

\begin{figure}[t]
 \begin{subfigure}[t]{0.49\textwidth}
  \centering
  \includegraphics[width=\textwidth]{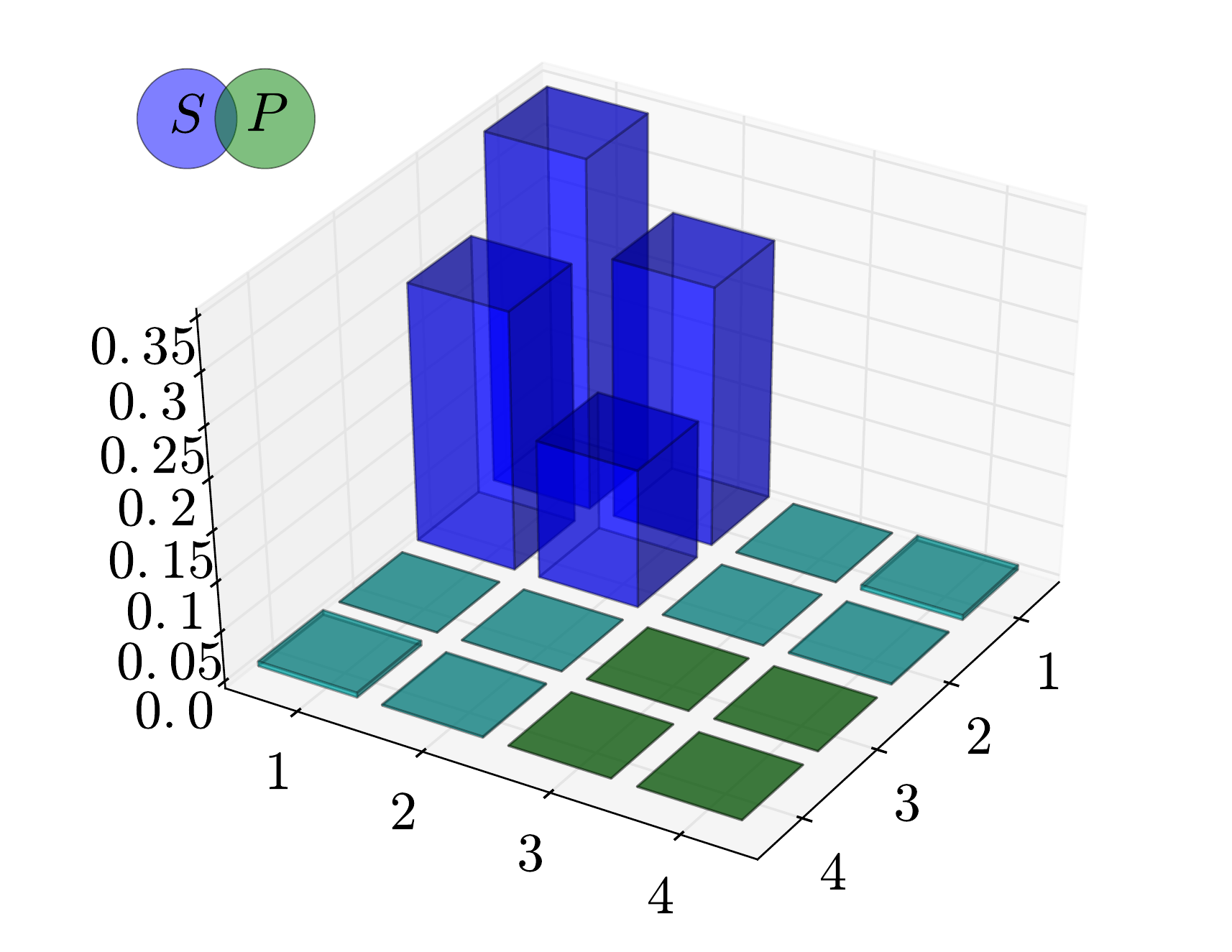}
    \caption{$0(0^{-+})$: $\eta_c$}		
 \end{subfigure}
 \begin{subfigure}[t]{0.49\textwidth}
  \centering
  \includegraphics[width=\textwidth]{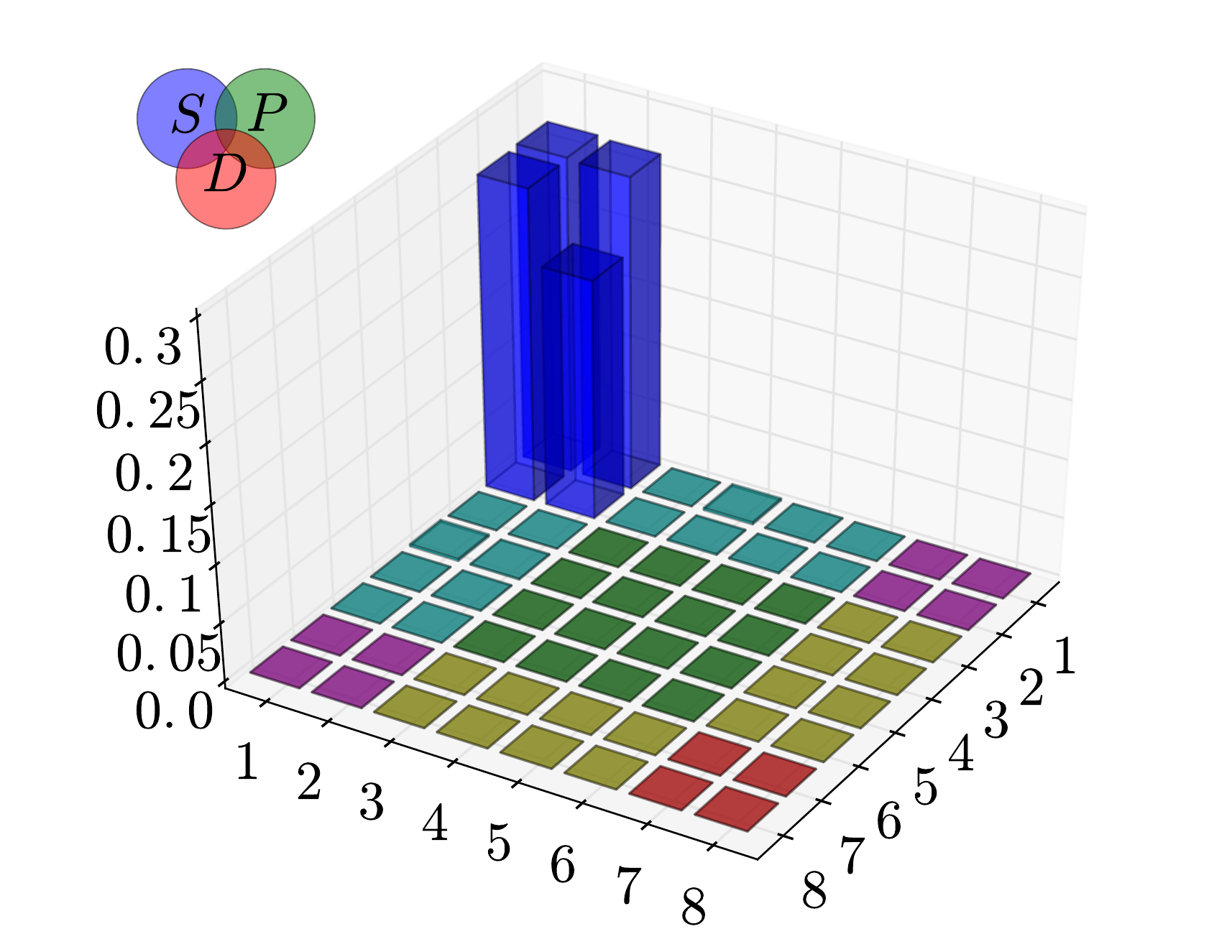}
    \caption{$0(1^{--})$: $J/\Psi$}		
 \end{subfigure}
 \begin{subfigure}[t]{0.49\textwidth}
  \centering
  \includegraphics[width=\textwidth]{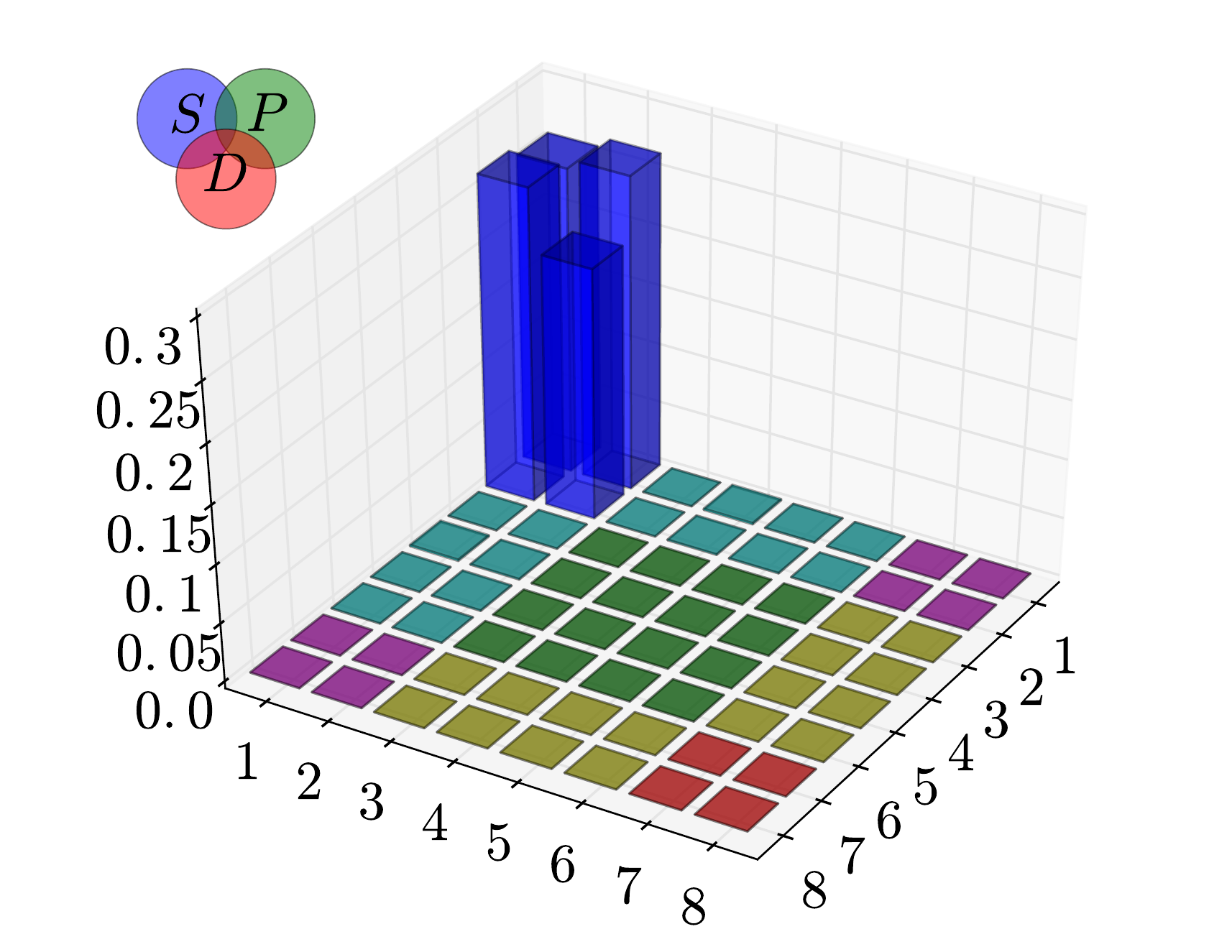}
    \caption{$1(1^{--})$: $\Psi(2S)$}		
 \end{subfigure}
 \begin{subfigure}[t]{0.49\textwidth}
  \centering
  \includegraphics[width=\textwidth]{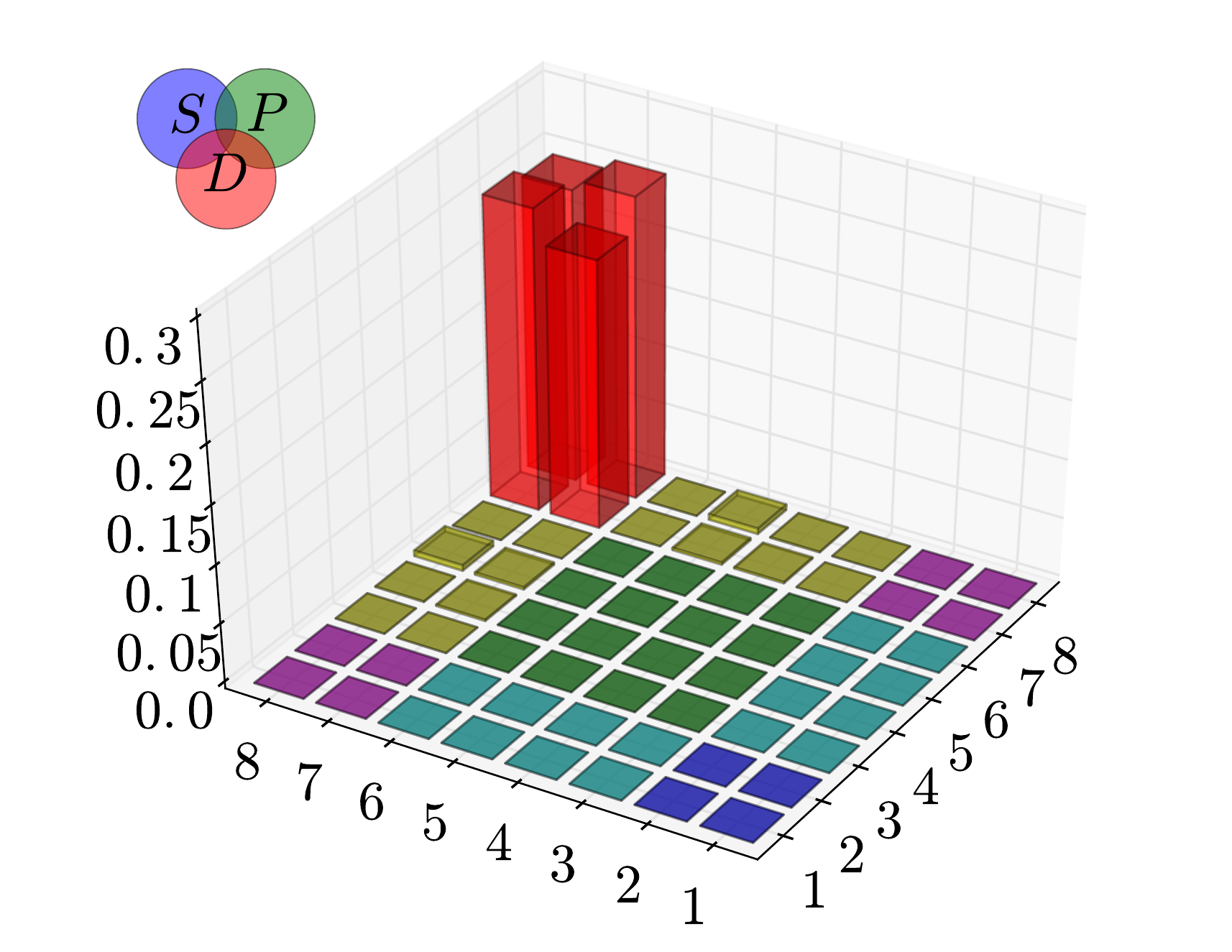}
    \caption{$2(1^{--})$: $\Psi(3770)$}		
 \end{subfigure}
 \begin{subfigure}[t]{0.49\textwidth}
  \centering
  \includegraphics[width=\textwidth]{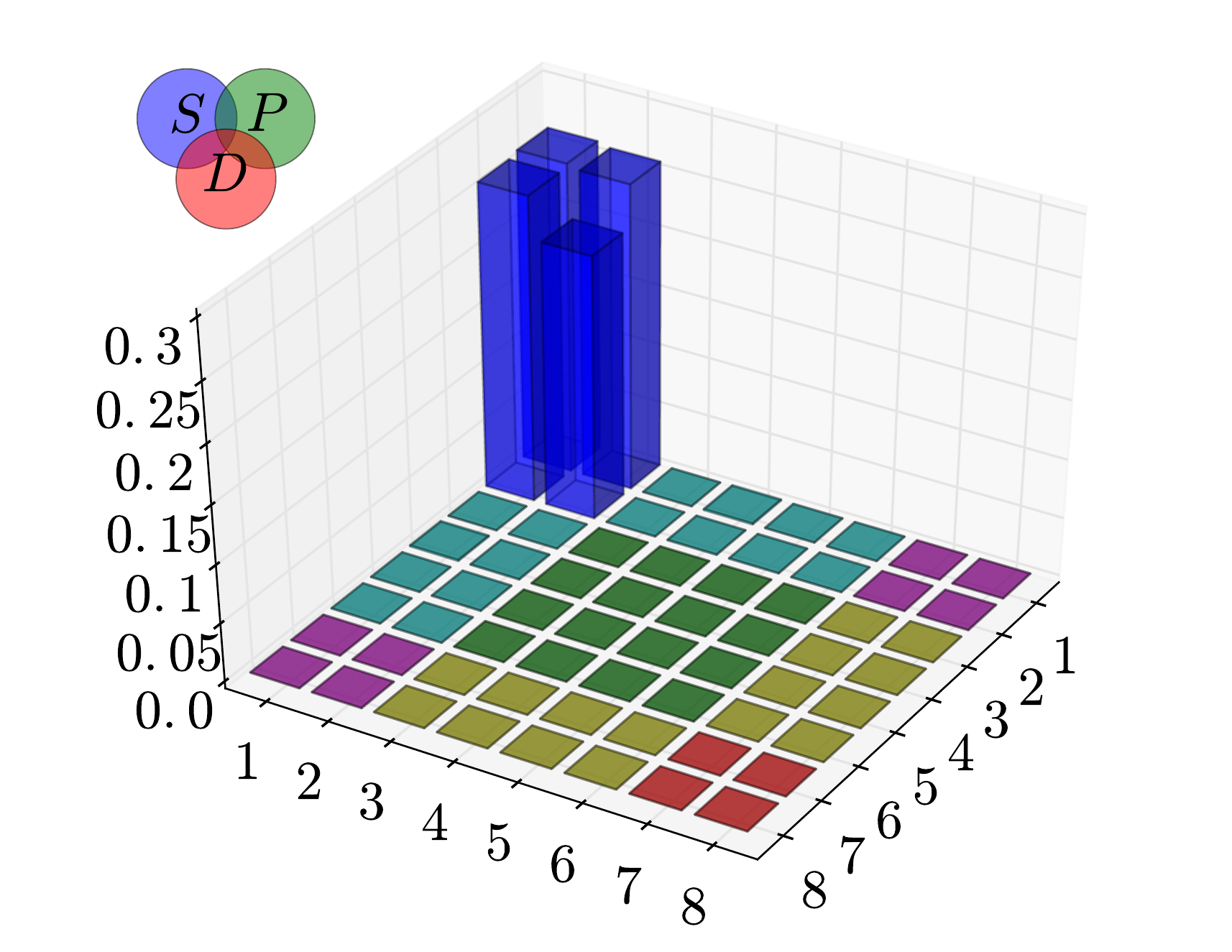}
    \caption{$3(1^{--})$: $\Psi(4040)$}		
 \end{subfigure}
 \begin{subfigure}[t]{0.49\textwidth}
  \centering
  \includegraphics[width=\textwidth]{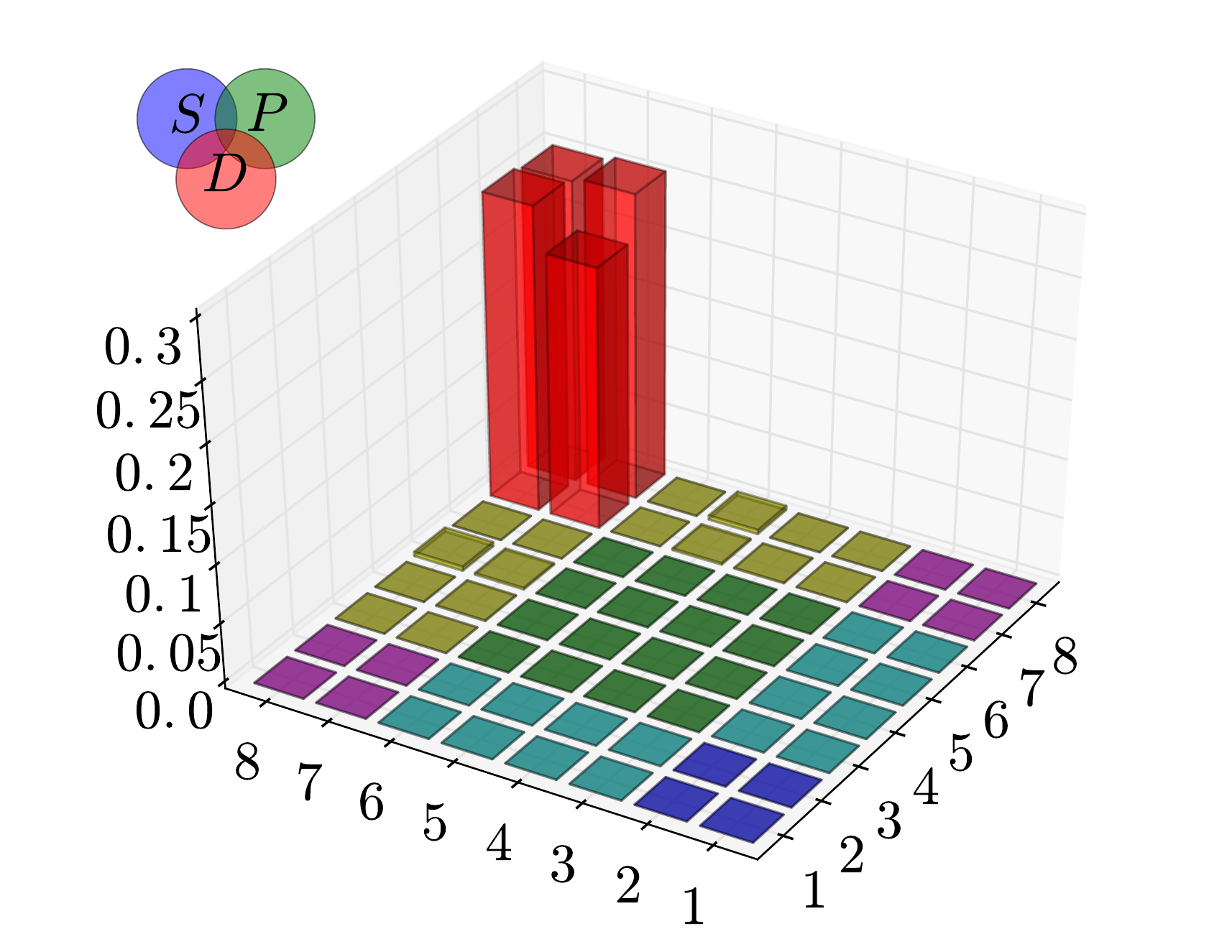}
    \caption{$4(1^{--})$: $\Psi(4160)$}		
 \end{subfigure}
\caption{\label{fig:oamd}Orbital angular momentum decompositions of charmonia. The set of quantum numbers underneath each subfigure denotes $n(J^{PC})$, where $n$ is the excitation quantum number, and $J$, $P$, and $C$ are the meson's total spin, parity, and charge-conjugation parity. }
\end{figure}

%%%%%%%%%
\section{DSBSE Model Setup}
In the DSBSE approach, chiral symmetry as well as its dynamical breaking are described by the axial-vector Ward-Takahashi identity (AVWTI) \cite{Munczek:1994zz}.
In any given truncation, satisfaction of the AVWTI leads to correct anchoring of results both in the absence \cite{Hilger:2015zva} and presence of D$\chi$SB \cite{Maris:1997tm}.
D$\chi$SB is already visible at the quark level by inspection of the solution of the quark Dyson-Schwinger equation (DSE), where the momentum dependent quark mass function is qualitatively different for the chirally symmetric and broken phases \cite{Blank:2010bz,Krassnigg:2016hml}.
At the meson level, one finds that, if a truncation satisfies the AVWTI, one automatically has a massless pion in the 
chiral limit, if chiral symmetry is dynamically broken, together with well-known features such as a generalized Gell-Mann-Oakes-Renner relation valid for all isovector pseudoscalar mesons irrespective of quark-mass content and level of  excitation \cite{Maris:1997tm,Holl:2004fr}.
In RL truncation, the quark-antiquark interaction is modeled by parameterizing an effective dressed one-gluon exchange as a function of the gluon momentum squared.
Several functional forms have been used for this purpose, and we chose the one in \cite{Maris:1999nt}, whose intermediate-momentum domain is characterized by two parameters, an inverse effective range $\omega$ and an overall strength $D$.
For more details and an illustration of this effective interaction, see Fig.~1 of \cite{Hilger:2015ora} and the related discussion there.

%%%%%%%%%%%%%%%%%%%%%%%%%%%%%%%%%%%%%%%%%%%%%%%%%%%%%%%%%%%%%%%%%%%%%%%%%%%%

\section{Orbital Angular Momentum}\label{sec:oamd}
The solution of the Bethe-Salpeter-equation (BSE) is the Bethe-Salpeter amplitude (BSA), which we obtain numerically, see \cite{Bhagwat:2007rj,Krassnigg:2008gd,Blank:2010bp,Blank:2010sn,Dorkin:2013rsa} for computational details. 
It is the covariant analog of a wave function and contains all information about the state \cite{Krassnigg:2010mh,Dorkin:2010ut,UweHilger:2012uua,Hilger:2015hka}. 
While being covariant, it also contains information that can be interpreted as the quark-antiquark orbital angular momentum in a meson by virtue of its covariants $T_i$, $i=1,\ldots,N$, where $N=4$ for spin-$0$ mesons or $N=8$ otherwise, and the index $i$ refers to the numbering on the horizontal axes in Fig.~\ref{fig:oamd}.
The details of the orbital angular momentum decomposition of a meson can be found in App.~A of \cite{Hilger:2015ora} and for baryons in \cite{Eichmann:2009zx}.
What is important to note here is that spin-$0$ allows for $S$-wave, $P$-wave and $S$-$P$ mixed components \cite{Bhagwat:2006xi}, while in a spin-$1$ meson BSA one can have $S$-wave, $P$-wave, $D$-wave,  and $S$-$P$, $P$-$D$, $S$-$D$ mixed components \cite{Hilger:2015ora}.
This situation as well as the quantification of these components via the canonical norm of the BSA  \cite{Smith:1969zk} are explained and illustrated in Fig.~6 of \cite{Hilger:2015ora} and the corresponding discussion there.

%%%%%%%%%
\section{Results and Discussion}\label{sec:results}
In previous studies we have focussed on heavy-quarkonium masses  \cite{Blank:2011ha,Hilger:2014nma} and leptonic decay constants \cite{Krassnigg:2016hml}.
In the latter work, we provided a likely assignment of $S$- and $D$-wave to vector quarkonia based on the calculated values for their leptonic decay constants, which we present again herein as $f_{\mathrm{Calc}}$ in Tab.~\ref{tab:dconstants}.
These numbers, as well as the orbital-angular-momentum assignment $L_{\mathrm{Calc}}$ presented  in the same table, were calculated with   $\omega=0.3$ GeV and $D=1.3$ GeV${}^2$ in the  interaction of \cite{Maris:1999nt}.
In this table, we compare them to the updated experimental values from \cite{Olive:2016xmw} denoted by $f_{\mathrm{Exp}}$.
In addition, we have computed the contributions of orbital angular momentum to the BSA for the states listed in Tab.~\ref{tab:dconstants}, where the percentages for each component fill the right part of the table. The corresponding illustrations can be found in Fig.~\ref{fig:oamd}. 

It is apparent that in all cases presented here the assignment of an overall or dominating quark-antiquark orbital angular momentum $L_{\mathrm{Calc}}$ is unambiguous. Our assignments confirm the statements made previously by means of $f_{\mathrm{Calc}}$ alone  \cite{Krassnigg:2016hml}.
Our leptonic-decay-constant and orbital-angular-momentum results are of interest in comparison to other recent approaches that have a similar or even more direct handle on such contributions in their respective meson amplitudes, e.\,g., \cite{Brambilla:2010cs,Dudek:2011bn,Li:2015zda,Rohrhofer:2016hiu,Leitao:2017esb,Leitao:2017bds,Piotrowska:2017rgt,Leitao:2017mlx}.

%%%%%%%%%
\ack
A.~K.~would like to thank the organizers of the FAIRNESS2017 workshop for their kind invitation, their hospitality in Sitges, 
and the once again stimulating and productive atmosphere.
This work was supported by the Austrian Science Fund (FWF) under project no.\ P25121-N27.

%%%%%%%%%
\section*{References}

%\bibliography{/Users/ank/Documents/library/had_nucl_graz}
\providecommand{\newblock}{}

\end{document}